\documentclass[prl,nofootinbib]{revtex4}
\usepackage{graphicx}
\usepackage{latexsym}
\def\be{\begin{equation}}
\def\ee{\end{equation}}
\def\bea{\begin{eqnarray}}
\def\eea{\end{eqnarray}}

\begin{document}
\title{The Interpretation of Inertial Force and The Existence of Absolute Background}

\author{ChiYi Chen${}^{a}$}
\email{chenchiyi@hznu.edu.cn}

\affiliation{${}^a$Hangzhou Normal University, Hangzhou 310036, China}

\begin{abstract}
The formalism of classical particle dynamics is reinvestigated according to the basic requirement of causal consistency, and a new equation of particle dynamics, which is more general and more in line with classical mechanics experiments than Newton's second law, is obtained. More importantly, the inertial frame of reference is no longer required and the inertial force is no longer introduced by hand. The new dynamical equation can be straightforwardly applied in any reference frame as long as which is irrotational with respect to the absolute background of the universe. The nature of the inertial force is nothing but the real forces acting on the reference object. The essence of this reform of classical particle dynamics is just a drawback of traditional formalism in Newtonian mechanics being corrected. However, it may be more important that a logical derivation of this new particle dynamics equation strongly suggests the existence of an absolute background for the whole space of the universe. In physical concepts, the absolute background for space does not conflict with the relative length of the base unit of space in Einstein's Special Relativity and General Relativity. Furthermore, the size of the base unit of space should essentially be understood as the length of the line segment, which is cut from the absolute background of space by the intrinsic physical events.
PACS number(s): 45.20.-d, 45.05.+x
\end{abstract}

\maketitle
\section{1 Introduction}
In the nonrelativistic framework of classical mechanics, the fundamental dynamics equation is Newton's second law.  But as is well known, Newton's second law is only valid in inertial reference frames. Provided that we apply the equation form of Newton's second law in a non-inertial reference frame, we need to introduce a fictitious force---inertial force by hand. The magnitude of the inertial force is usually determined by the relative acceleration between the non-inertial reference frame in question and a certain inertial reference frame\cite{Shouzhu,Douglas,Feynman}. Therefore, the classical particle dynamics is totally based on the concept of inertial frame of reference. However, as well-known to us, we are never able to find a real inertial reference frame in practice. This situation is surely not satisfactory\cite{Machprinciple,Liuliao}. On the other hand, there is a pending question which exists in astronomical observation, namely why the orbital velocities of stars in the Milky Way galaxy is greater than the prediction calculated from the "luminous matter" it contains? In a most popular approach it is explained from the point of view of dark matter\cite{darkmatter}. Besides, a theory of modified Newtonian dynamics($MOND$) was proposed by Mordehai Milgrom in 1983. In this theory, the problem of flat rotation curve was explained by modifying the dynamics of classical mechanics, or modifying Newton's gravity theory. Milgrom found that the flat rotation curve problem can be elegantly solved if the Newtonian dynamics is modified only when the acceleration of reference frames is very small\cite{milgrom}. However, from recent results of laboratory tests, the approach such as $MOND$ theory to modify Newtonian dynamics by artificially introducing a scaling factor is not so promising\cite{ignatiev,gundlach}. Therefore, in this paper we reinvestigate the reasonability of the formulism in nonrelativistic classical dynamics, starting from the most fundamental requirement---the principle of causal consistency.

In this paper, we explore the reasonable formalism for particle dynamics under the nonrelativistic framework of classical mechanics. The paper is organized as follows. In Sec.2, we present the traditional formalism of particle dynamics. A causal inconsistency problem which exists in the theoretical formula of Newton's second law is pointed out. In Sec.3, we reconstruct a new formalism of particle dynamics under the guidance of causal consistency principle. In Sec.4, the new particle dynamical equation is examined by making a comparison with the empirical laws from classical mechanics experiments. We show that the new dynamical equation is able to provide a more natural picture to explain all these empirical laws than Newton's second law. At the same time, the inertial frame of reference is no longer required and the inertial force is no longer introduced by hand.
In Sec.5, it is fully shown that the new classical particle dynamics equation (9) is an indispensable development of Newton's second law if Newton's second law is restored to be an empirical law, which is essentially a quasi-differential causal relationship for particle dynamics.
In Sec.6, we discuss the physical significance of the new dynamical equation. In the appendix, we propose a new fundamental physical picture for spacetime since the new dynamical equation (9) must be valid under the nonrelativistic framework of classical mechanics, and the logical derivation of this equation strongly suggests the existence of an absolute background for the whole space of the universe.

\section{2 formalism of Newtonian Particle Dynamics and Existing Problems}
In Newtonian mechanics, the main equation of particle dynamics is given by Newton's second law\cite{Douglas,Feynman}. The mathematical form can be generally expressed as
\begin {eqnarray}
{\bf F}|_{p}=m_{p}{\bf a}|_{p-O}.
\end {eqnarray}
In theory, the left hand side of this equation must denote the total force from the whole universe acting on the particle $p$. Otherwise, when the equation is applied into concrete cases, we will not be able to make it clear what forces should be included in the count, and what forces should not be counted. The $m_{p}$ of the right hand side of this equation denotes the mass of the particle $p$; ${\bf a}|_{p-O}$ denotes the acceleration of particle $p$ with respect to the reference frame $O$. Here the real reference object which is assigned to define this reference frame $O$ may as well be denoted by $O$. Since the reference object is definitely fixed in the reference frame, the acceleration (${\bf a}|_{p-O}$) of the particle $p$  with respect to the reference frame $O$, is equivalently measured relative to the reference object $O$.

It is well known to us that Newton's second law (1) is valid only for inertial reference frames. But by now we have not found even one exact inertial reference frame in practice. Actually, it is also not possible for us to find an exact inertial reference frame, since all real matter in the universe are acted on by the gravitational forces from other matter and so are in a never-ceasing motion and change. Therefore, any real reference frame must be a non-inertial reference frame. According to Newtonian mechanics, if we want to make a mechanical application of Newton's second law in a non-inertial reference frame, in theory we still need to find out an inertial reference frame firstly. After that, we can introduce a fictitious inertial force according to the relative acceleration between the inertial and non-inertial reference frames\cite{Machprinciple,inertialforce},
\begin {eqnarray}
{\bf f}|_{inertial}=-m_{p}{\bf a}|_{O'-O}.
\end {eqnarray}
Therefore, in the application of Newton's second law, it is always required to approximate a real reference frame to be a inertial reference frame. It should be noted that such an approximation is actually made at the stage of theory application, but not at the stage of the practical measurement of physical quantities. In this sense, Newtonian mechanics is still not a perfect theory.

In fact, most of classical mechanics experiments are conducted and analyzed under the ground-based laboratory reference frame. If we assume the object under study is denoted by $A$, and the ground-based laboratory reference frame is denoted by $B$, the empirical law which is satisfied by all these ground-based classical mechanics experiments can be written down as,
\begin {eqnarray}
({\bf f}_{A})_{NonGrav}+m_{A}{\bf g}=m_{A}{\bf a}|_{A-B}.
\end {eqnarray}
Here ${\bf g}$ is the gravitational acceleration on the ground. It should be noted that we don't need to count the gravitational forces from the outside of the earth system. This is far from being negligible since the orbit acceleration of the earth around the sun reaches to $6\times10^{-3}m/s^{2}$. Besides, another empirical law well known to us is that at the planetary scale. Here the sun-centered reference frame is usually favored, and the empirical law which is satisfied by the mechanical experiments at the solar system is given by
\begin {eqnarray}
({\bf f_{1}})_{SolarGrav}=m_{1}{\bf a}|_{1-2}.
\end {eqnarray}
Here $1$ denotes the celestial body under study, and $2$ denotes the sun-centered reference frame. $({\bf f_{1}})_{SolarGrav}$ denotes the gravitational forces acting on the particle $1$ from the matter in solar system, rather than the total forces exerted by the whole universe.

In the practical application, both the equations (3) and (4) have achieved a remarkable success in precision. But if we make a comparison between equations (3-4) and theoretical formula of Newton's second law (1), we can find that the reference frames actually used in (3) and (4) are not the inertial reference frame which is compulsory required by (1). At the same time, their forces being counted are not the total forces which is also theoretically required by Newton's second law (1). In fact, applying Newton's second law to explain above successful empirical laws is a complex and subtle thing\cite{Machprinciple}. Roughly, it can be understood according to the following steps. First, a gravitationally bounding system at a more large scale should be adopted as an approximated inertial reference frame, meanwhile the counting of the forces should also be limited to that exerted only by the matter inside of this more large scale system. According to the relative acceleration between this approximated inertial reference frame and the actually used reference frame under study, the inertial force in this case can be calculated. Second, since this actually used reference frame is not an exact inertial reference frame, this inertial force should be added into the equation (1) by hand so we can approximately obtain empirical laws like (3) and (4). However, if we want to further explain why the counting of the forces can be limited only from above more large system, we must invoke above steps once again\cite{Machprinciple}. Therefore, the Newton's second law, as a theoretical formula, fails to give a concise and elegant physical picture in the interpretation of empirical laws of classical mechanics experiments.

\section{3 Causal Consistency Principle and New formalism of Particle Dynamics}
Since for Newton's second law, neither the current theoretical formula nor the physical picture is satisfactory, we consider whether it is possible to reconstruct the physical logic of particle dynamics. In this process, the only one most fundamental principle which can be resorted to is the causal consistency principle.

First of all, every reference frame must be established on a real reference object. Otherwise, there would be no reference value in measuring any object's motion in the natural world. Therefore, a physical reference frame must be the real reference frame. As for the relationship between the reference object and the reference frame, the reference frame can be naturally established by identifying the reference object as its origin point if we assign a reference object first in practical cases. Otherwise, in principle any real object which is fixed in the reference frame can be identified as its reference object if the reference frame is assigned first. For example, as a most usual reference frame, the ground-based reference frame actually selects an arbitrary object fixed on the ground as the reference object.

Essentially, Newton's second law should be a causal law of particle dynamics. Here the forces acting on the particle under study should be the cause and the resulting acceleration should be the effect. In the traditional theoretical formula of Newton's second law : ${\bf F}|_{p}=m_{p}{\bf a}|_{p-O}$, the left hand side ${\bf F}|_{p}$ is the total force acting on the particle $p$ from the whole universe, which only depends on $p$. Yet the right hand side ${\bf a}|_{p-O}$ is the acceleration of the particle $p$ with respect to the reference frame $O$, equivalently measured relative to the reference object $O$. Therefore in fact, the effect (namely the result) ${\bf a}|_{p-O}$ depends not only on the particle $p$, but also on the reference object $O$ which corresponds to the origin point of the reference frame. In this sense, the causality of Newton's second law is not symmetric and consistent in the form. If the current theoretical formula of Newton's second law is desired to be directly applicable in any practical reference frames, it is not reasonable for that only the motion of the reference object gets involved into the current theoretical formula of Newton's second law, but the corresponding force acting on the reference object itself is never under consideration, since any practical reference frames must be defined according to real reference objects. In other words, Newton's second law has a problem of causal inconsistency. This is the very point to account for why Newton's second law is theoretically valid only in inertial reference frames, but none of them can be found in practice.

Then how to solve the problem of causal inconsistency? The key point is how to describe the corresponding effect according to the principle of causal consistency, if the total force from the whole universe acting on the particle is considered. We still suppose that the particle dynamics is certainly to be a theory with the principle of causal consistency. We regard forces as the cause, and regard accelerations as the effect(result). The total force acting on a single particle should be objective, namely it will not change with the variation of observers. Therefore, the corresponding effect should also be objective, and not relevant with any reference frame. In this way, a completely objective acceleration can only be expressed as the acceleration with respect to the absolute background of the whole universe,
\begin {eqnarray}
{\bf F}|_{p}= m_{p}\frac{d^{2}}{dt^{2}}\bigodot|_{p}.
\end {eqnarray}
Here the objective position of the particle $p$ in the absolute background of space is particularly denoted by $\bigodot|_{p}$. In concepts we should distinguish the relative length or duration of base units of space-time from the absolute background of space-time. Just as its name implies, the base unit of space is the spatial span of the standard one meter(namely the standard ruler), and the base unit of time is the time span of the standard one second(namely the standard clock). The base units of space-time can be changed according to the laws of Einstein's special relativity and geometric theory of gravity. But the background for space-time must exist as the absolute basis(or reference background) to reflect this change of base units.

It is worth stressing that, the absolute background for space can completely understood as the inherent part of Newton's absolute view of space-time, since the latter is a natural philosophy basis for classical mechanics. Therefore, to say the least, in above derivation we are also able to regard $\bigodot|_{p}$ as the objective position of the particle $p$ in the absolute space. As far as the framework of classical mechanics is concerned, it has no objections in logic if here we directly invoke the physical concept of Newton's absolute space-time. However from the point of view of doing science, it would be better to keep assumptions as less as possible in every step of the derivation for any verifiable physics law. Because only in this way, a most clear and direct causal structure can be reached for the whole theory. Therefore, in this paper we propose a compromise on space-time views. The main idea is that the concept of
space-time should be further subdivided into two levels(or two aspects). One is the relative length or duration of base units of space-time, and the other is the absolute background for space-time. The relative length or duration of base units of space-time can be regarded as a unit line segment which is cut from the absolute background for space-time.

An absolute background existing for the whole space of our universe is firstly originated from our intuitional experiences. The absolute background of space (also be called cosmic spatial background) can be intuitively understood. As the name implies, the spatial background is just what still exist in a space region after all objects inside it were moved away. And the cosmic spatial background is just what still exist in the whole universe after all concrete objects in the universe were moved away. Here "absolute" means that the background of space as the most fundamental reference is defined to be flat, homogeneous and infinite. Besides, in order to describe any object's state of motion in the universe, the background of the whole space in our universe as the reference should be independent of the motion and distribution of all objects which exist in the universe. There is no concept of base unit for the background of space-time since the background itself does not contain any specific object. On the contrary, only the base units of space-time are defined by local intrinsic events which occur in real objects. For this reason, we must distinguish in physics between the absolute background and the relative base units of space.

It is natural that the motion of all objects in the universe must be performed over this common absolute three-dimensional space background, because only on this basis then the existence of an objective dynamical law is possible. From the logical point of view, every natural principle describing a pure relative law must have an absolute basis. There is one point should be further emphasized. Here the absoluteness is proposed only for the background of space-time, instead of base units of space-time. This is key point to keep the absolute background of space conceptually compatible with the solidest part of modern physics. What Einstein's special theory of relativity has actually denied is only that the base units of space or time are absolute. What Einstein's general theory of relativity has further proved is that the base units of space or time will change according to the gravitational field\cite{Sanhui,weinberg}. All the dynamical laws in these theories does in conflict with the absolute background of space-time. Moreover, in the logic system of special relativity, it is easy to see that every event is assumed to have an objective position\cite{Sanhui}. In essence, this is just the reflection of the existence of an absolute background for the universe. Only based on the absolute background of space, it is possible for us to assume that any particle at any time has its objective position in the universe. We conjecture that both the number of types of interactions in the universe and their calculation rules are objective. In principle, all kinds of interactions can be recognized and understood by people, and their calculation rules can be ultimately obtained. The reason is just that all these interactions must be able to interpret the motions of all objects in the universe, simultaneously and self-consistently.

To say the least, the existence of the absolute background of space itself is the underlying part of Newton's absolute view of space-time under the framework of classical mechanics. Fortunately, this underlying part is still not denied by Einstein's special theory of relativity. Therefore, in this paper it always be rational for us to discuss the formalism of particle dynamics based on the absolute background of space. Although every particle has its objective position in the absolute background of space, there is still a problem that the objective position in the absolute background of space can not be directly measured. What we can really measure is the difference between any two objective positions, which substantially constructs a mathematical vector,
\begin {eqnarray}
{\bf r}|_{p-O}= \bigodot|_{p}- \bigodot|_{O}.
\end {eqnarray}
After that, we are able to construct a particle dynamical equation which is really available to practical observers. In fact, every reference frame must be established on a real reference object. Otherwise, there would be no reference value in measuring any object's motion in the natural world. Therefore, a physical reference frame must be the real reference frame. As for the relationship between the reference object and the reference frame, the reference frame can be naturally established by identifying the reference object as its origin point if we assign a reference object first in practical cases. Otherwise, in principle any real object which is fixed in the reference frame can be identified as its reference object if the reference frame is assigned first. All objects in the universe, including objects under study ($p$) and reference objects ($O$), should be of equal status in the most fundamental law of dynamics. Similarly, the dynamics of any real reference object should also satisfy
\begin {eqnarray}
{\bf F}|_{O}=m_{O}\frac{d^{2}}{dt^{2}}\bigodot|_{O}.
\end {eqnarray}
Here the reference object $O$ naturally corresponds to the origin point of a reference frame, so we can establish a reference frame which is irrotational with respect to the absolute background of space. The introduction of reference frames is just to make relative measurements on kinematical quantities. As a causal correspondence, the forces should also be relatively counted in nature.
\begin {eqnarray}
m_{O}{\bf F}|_{p}-m_{p}{\bf F}|_{O} = m_{p}m_{O}\frac{d^{2}}{dt^{2}}( \bigodot|_{p}- \bigodot|_{O})= m_{p}m_{O}\frac{d^{2}{\bf r}|_{p-O}}{dt^{2}}.
\end {eqnarray}
Finally, we obtain
\begin {eqnarray}
\frac{{\bf F}|_{p}}{m_{p}}-\frac{{\bf F}|_{O}}{m_{O}}={\bf a}|_{p-O}.
\end {eqnarray}
In this equation, the definition of the force and the acceleration are just the same as that in the traditional theoretical formula of Newton's second law (1). ${\bf F}|_{p}$ and ${\bf F}|_{O}$ are the total forces from the whole universe acting on the particle $p$ and the reference object $O$ respectively. $m_{p}$ and $m_{O}$ denote the mass of the particle $p$ and the reference object $O$ respectively. Since the reference object is definitely fixed in the reference frame, the acceleration (${\bf a}|_{p-O}$) of the particle $p$ is actually measured relative to the reference object, equivalently with respect to the origin point of the reference frame. Keep in mind, any physical reference frame must be related to a certain real reference object. Therefore, the equation (9) is just the dynamical equation of the particle $p$ with respect to the reference frame $O$.

In the equation (9), the object under study and the reference object are now placed on an equal status. So the special status for the reference object in the universe is removed. It reflects that all objects in the universe have the equal status in dynamics. That is obvious as well, since what object should be selected as the particle under study, and what particle should be selected as the reference object are essentially assigned by people. In fact, there is no essential division of them. More importantly, the nature of the inertial force is nothing but the real force acting on the reference object, and which is supposed to appear in the new dynamical equation (9) according to the principle of causal consistency. To demonstrate the difference between the equation (9) and the theoretical formula of Newton's second law (1), we may rewrite (9) to be,
\begin {eqnarray}
 {\bf F}|_{p}-\frac{m_{p}}{m_{O}}{\bf F}|_{O}=m_{p}{\bf a}|_{p-O}.
\end {eqnarray}
Here the left hand side of this equation can be called as a relative counting of forces. Obviously, the equation (10) has a net term ($-\frac{m_{p}}{m_{O}}{\bf F}|_{O}$) more than Newton's second law, while the other terms are identical. if the new particle dynamical equation (10) takes a special case: ${\bf F}|_{O}=0$, namely the case that the total forces acting on the reference object equals to zero, above formula goes back to Newton's second law. Therefore, in fact, the theoretical formula of Newton's second law can be derived from the new particle dynamics equation (10) as an extreme case. This is the key point. The equations (9) and (10) are just the new dynamical equation which is proposed to replace the current theoretical formula of Newton's second law under the framework of classical mechanics, since the new dynamical law (9) or (10) presents a more natural and concise physical picture based on reinterpreting all empirical laws from classical mechanics experiments. In the application of (9) or (10), it is easy to find that the inertial reference frame is no longer required and the inertial force is no longer introduced by hand.

\section{4 Practical Examination of the New Particle Dynamical Equation }
\subsection{4.1 the realization of general principle of relativity and the nature of inertial force}
The equation (10) can be directly applied in any reference frame which is irrotational with respect to the absolute background of space, and naturally interprets the inertial force. In other words, a quasi-general principle of relativity is realized. In the last paragraph of Sec.5, it will be illustrated that such a general principle of relativity is more realistic than Einstein's general principle of relativity. Obviously, ${\bf F}|_{p}$ and ${\bf F}|_{O}$ in the equation (10) can be calculated respectively according to existing knowledge of all types of interactions. In principle any real object or arbitrary part of an existing object can be identified as a reference object. Therefore, the equation (10) can be applied in all irrotational reference frames. Moreover, according to the equation (10), the inertial force is essentially the real force acting on the reference object, which must be deducted in a relative counting of forces. To illustrate this point, we assume that there are two relatively accelerated reference frames. Their reference objects are respectively denoted as $O$(assumed to be the reference object of a so-called inertial reference frame) and $O'$( of a real reference frame in practice). For above two reference frames, it is the difference in the forces acting on the reference objects that results in a relative acceleration between these two reference frames. In Newtonian mechanics, such a relative acceleration is depicted by a so-called inertial force. But now the relative acceleration between reference frames is naturally interpreted by the difference in the forces acting on their reference objects. Recalling the definition of the inertial force in the equation (2), for any real reference frame $O'$ which is certainly a non-inertial reference frame, we always have
\begin {eqnarray}
{\bf f}|_{inertial}=-m_{p}{\bf a}|_{O'-O}=m_{p}{\bf a}|_{p-O'}-m_{p}{\bf a}|_{p-O}=-\frac{m_{p}}{m_{O'}}{\bf F}|_{O'}+\frac{m_{p}}{m_{O}}{\bf F}|_{O}=-\frac{m_{p}}{m_{O'}}{\bf F}|_{O'}.
\end {eqnarray}
Here ${\bf F}|_{O'}$ is the total forces acting on the real reference object $O'$ from the whole universe. Above analysis illustrates that the new dynamical equation (9)(or (10)), compared with Newton's second law, has been endowed with a more complete and symmetric theoretical structure and so possesses the real universality. More importantly, in the application of (9)(or (10)), the inertial force is no longer introduced by hand.

\subsection{4.2 The validity of the new dynamical equation}
It can be proved that all successful empirical laws from classical mechanics experiments can also be naturally accounted for by the new dynamical equation (10). In general, we assume two arbitrary objects from the solar system denoted by 1 and 2(such as the Moon and Venus). The total force from the whole universe acting on the particle $i$ can be written down as (Here we assume that non-gravitational forces are negligible)
 \begin {eqnarray}
{\bf F}|_{i}=({\bf f}_{i})_{NonGrav}+({\bf f}_{i})_{SolarGrav}+({\bf f}_{i})_{OutSolarGrav}\approx ({\bf f}_{i})_{SolarGrav}+({\bf f}_{i})_{OutSolarGrav}.
\end {eqnarray}
Here $({\bf f}_{i})_{SolarGrav}$ means the gravitational forces acting on the particle $i$ from the matter in solar system, and $({\bf f}_{i})_{OutSolarGrav}$ the gravitational forces acting on the particle $i$ from the matter outside of the solar system. According to the equations (10) and (12), the relative dynamics between 1 and 2 can be expanded as
\begin {eqnarray}
{\bf F}|_{1}-\frac{m_{1}}{m_{2}}{\bf F}|_{2}\simeq[({\bf f_{1}})_{SolarGrav}-\frac{m_{1}}{m_{2}}({\bf f_{2}})_{SolarGrav}]+[({\bf f_{1}})_{OutSolarGrav}-\frac{m_{1}}{m_{2}}({\bf f_{2}})_{OutSolarGrav}]=m_{1}{\bf a}|_{1-2}.
\end {eqnarray}
Owing to both 1 and 2 being bound in the same solar system, the gravitational forces exerted by the matter outside of the solar system satisfy,
 \begin {eqnarray}
\frac{({\bf f_{1}})_{OutSolarGrav}}{({\bf f_{2}})_{OutSolarGrav}}\approx\frac{m_{1}}{m_{2}}.
\end {eqnarray}
So the relative dynamics (13) between objects 1 and 2 can be approximately rewritten as
  \begin {eqnarray}
{\bf F}|_{1}-\frac{m_{1}}{m_{2}}{\bf F}|_{2}\approx[({\bf f_{1}})_{SolarGrav}-\frac{m_{1}}{m_{2}}({\bf f_{2}})_{SolarGrav}]=m_{1}{\bf a}|_{1-2}.
\end {eqnarray}
Specifically, if we select the center of sun as the origin point (the location of particle 2) of the reference frame in the above equation, at that time the reference object 2 can be the whole sun, or can be any real object located at the center of sun. For instance, if we choose the whole sun as the reference object, the second term in the left hand side of the above equation would be suppressed since $m_{1}\ll m_{2}$. Otherwise, we may choose a real object at the center of sun as the reference object 2, which has a comparable mass as the object 1. Since for the reference object 2, the distribution of the matter in the solar system has a high approximated spherical symmetry, the forces acting on the reference object 2 from the solar system may usually be approximated to be zero: $({\bf f_{2}})_{SolarGrav}\approx0$. Therefore, if we don't desire a high precision, the dynamics of object 1 can always be expressed as
\begin {eqnarray}
{\bf F}|_{1}-\frac{m_{1}}{m_{2}}{\bf F}|_{2}\approx({\bf f_{1}})_{SolarGrav}=m_{1}{\bf a}|_{1-2}.
\end {eqnarray}
That is the empirical law well satisfied by celestial bodies in the solar system, namely the equation (4) is recovered. Obviously, above empirical law is substantially an approximation of the new dynamical equation (10) when it is applied in gravitational bound systems (It undergoes about two steps of approximations). The approximation such as (16) is valid for usual sun-centered reference frame, earth-centered reference frame and innumerable other gravitational bound systems in our universe. Even for the most common ground-based reference frame, it substantially comes down to the earth-centered reference frame. Therefore, all successful empirical laws from classical mechanics experiments can be naturally accounted for again by the new dynamical equation (10). Besides, why the gravitational forces exerted by the outside matter of the solar system do not have to be taken into account? The direct reason is not because the magnitude of these forces is so small that all of them can be ignored, but owing to the fact that these forces are deducted in the relative counting of forces.

Now we consider the case that the new dynamical equation (10) is applied in ground-based reference frame. We assume that $A$ is the particle under consideration which is moving on the ground and $B$ denotes a frame of reference which is irrotational with respect to the background of the universe and its reference object is at rest with respect to to the earth's surface. According to the equation (10), the relative dynamics between $A$ and $B$ can be directly expressed as
\begin {eqnarray}
{\bf F}|_{A}-\frac{m_{A}}{m_{B}}{\bf F}|_{B}=m_{A}{\bf a}|_{A-B}.
\end {eqnarray}
In theory, the total forces acting on the $A$ and $B$ can be expanded as
\begin {eqnarray}
{\bf F}|_{A}=({\bf f}_{A})_{OutEarthGrav}(\propto m_{A})+({\bf f}_{A})_{EarthGrav}(\propto m_{A})+({\bf f}_{A})_{NonGrav},\cr
{\bf F}|_{B}=({\bf f}_{B})_{OutEarthGrav}(\propto m_{B})+({\bf f}_{B})_{EarthGrav}(\propto m_{B})+({\bf f}_{B})_{NonGrav}.
\end {eqnarray}
The non-gravitational forces acting on the reference object $B$ can be calculated from an equivalent case that the object $B$ hanging under a gravity measuring instrument located at the same place remains still, since both of them are in the static force balance. In that case, the non-gravitational force must be equal to the pulling force from the instrument,
\begin {eqnarray}
({\bf f}_{B})_{NonGrav}=-m_{B}{\bf g}.
\end {eqnarray}
Substituting it into the equation (17), at the same time we adopting an approximation that the gravitational forces from the earth and outside of the earth can all be counterbalanced between A and B, we finally obtain,
\begin {eqnarray}
{\bf F}|_{A}-\frac{m_{A}}{m_{B}}{\bf F}|_{B}&=&({\bf f}_{A})_{NonGrav}-\frac{m_{A}}{m_{B}}(({\bf f}_{B})_{NonGrav})=({\bf f}_{A})_{NonGrav}-\frac{m_{A}}{m_{B}}(-m_{B}{\bf g})\cr
&=&({\bf f}_{A})_{NonGrav}+m_{A}{\bf g}=m_{A}{\bf a}|_{A-B}.
\end {eqnarray}
This is the empirical law well satisfied by the classical mechanics experiment when it is conducted and analyzed under ground-based reference frames, namely the equation (3) is recovered. There is one point which should be emphasized. The change from $({\bf f}_{A})_{EarthGrav}$ in (18) to $m_{A}{\bf g}$ in (20), is usually interpreted by a fictitious inertial force owing to the rotation of the earth. But now we can see that it actually should be attributed to the real force acting on the reference object $B$ from the ground. Broadly speaking, the surface of the earth is rigid in spatial distance relative to the center of the earth, so the measurement of kinematical effect in ground-based reference frames is equivalent to that relative to the center of earth if the rotation of the earth is not taken into account. Accordingly, for the part of counted force, the gravitational forces exerted by the outside matter of the earth system should not be considered.

Furthermore, if we assume the reference object $B$ an arbitrary object moving over the ground(including the case of acceleration), at that time $({\bf f}_{B})_{NonGrav}\neq-m_{B}{\bf g}$, so the formula of particle dynamical equation (20) should be changed into,
\begin {eqnarray}
({\bf f}_{A})_{NonGrav}-\frac{m_{A}}{m_{B}}({\bf f}_{B})_{NonGrav}=m_{A}{\bf a}|_{A-B}=[({\bf f}_{A})_{NonGrav}+m_{A}{\bf g}]-\frac{m_{A}}{m_{B}}[({\bf f}_{B})_{NonGrav}+m_{B}{\bf g}].
\end {eqnarray}
Therefore, in all senses, the empirical laws from classical mechanics experiments can always be regarded as the approximation of the new dynamical equation (9) (or (10)) under some special conditions.

\subsection{4.3 the universality of the new dynamical equation}
The equation (10) is directly applicable in the relative dynamics between galaxy clusters. For instance, we assume that there are only two particles (such as $1$ and $2$) existing in the whole universe, and there is only gravitational interaction between them. In the framework of Newtonian mechanics, none of real particles can be approximated as an inertial reference frame, so Newton's second law can not be directly applied in this situation. Even someone resorts to the center of mass for this system, but essentially the center of mass method can be finally attributed to an assumption that the center of mass is at rest with respect to a certain inertial reference frame. However, according to the equation (10) the relative dynamics between above two particles can be directly written down as
  \begin {eqnarray}
{\bf F}|_{1}-\frac{m_{1}}{m_{2}}{\bf F}|_{2}&=&(\frac{Gm_{1}m_{2}}{r^{3}}{\bf r}_{1\rightarrow 2})-\frac{m_{1}}{m_{2}}(\frac{Gm_{1}m_{2}}{r^{3}}{\bf r}_{2\rightarrow 1})=m_{1}{\bf a}|_{1-2}=m_{1}\frac{d^{2}{\bf r}_{2\rightarrow 1}}{dt^{2}}.
\end {eqnarray}
This case also shows that in the application of the new dynamical equation (9) (or (10)), the inertial reference frame is no longer required.

\section{5 the Relationship between the New Dynamical Equation and Newton's Second Law}
By now we have seen that, the new dynamical equation (9) naturally accords with existing all classical mechanics experiments. In the meantime, the new dynamical equation (9) is more universal than Newton's second law. Especially, the theoretical necessity of the existence of inertial reference frame and inertial force is logically removed from the formalism of particle dynamics. But more importantly, the new dynamical equation (9) is more in line with the causal consistency principle. Even so, there is one thing should still be pointed out that the new dynamical equation (9) can be exactly derived under the framework of Newtonian mechanics. Since in all concepts of reference frame, besides the property of space-time base units, there is only the reference object is physical. Therefore under the framework of classical mechanics, any object's dynamical problem in any reference frame can be substantially regarded as a two-body problem, if the space-time base units have been assumed to be constant for all reference frames. Now we derive the dynamical equation based on Newtonian mechanics to solve such a two-body problem. Firstly, we assume there really exists an inertial reference frame which is denoted by $\Omega$. According to the theoretical formula of Newton's second law, an arbitrary object $p$ satisfies,
\begin {eqnarray}
{\bf F}|_{p}=m_{p}{\bf a}|_{p-\Omega}.
\end {eqnarray}
Here, the forces ${\bf F}|_{p}$ is defined to include all the forces acting on the particle $p$ from the whole universe. Similarly, to an arbitrary reference object $O$, which must also obey the same natural law,
\begin {eqnarray}
{\bf F}|_{O}=m_{O}{\bf a}|_{O-\Omega}.
\end {eqnarray}
Performing a simple algebraic manipulation, we obtain
\begin {eqnarray}
{\bf F}|_{p}-\frac{m_{p}}{m_{O}}{\bf F}|_{O}=m_{p}{\bf a}|_{p-\Omega}-\frac{m_{p}}{m_{O}}(m_{O}{\bf a}|_{O-\Omega})=
m_{p}[{\bf a}|_{p-\Omega}-{\bf a}|_{O-\Omega}]= m_{p}{\bf a}|_{p-O}.
\end {eqnarray}
Therefore, under the framework of Newtonian mechanics, the new dynamical equation (10) is recovered. Why the new form of particle dynamical equation was not explored and not paid enough attention in past several centuries, I think a possible reason is that people have never recognized that the dynamical dependence on reference frames can be finally attributed to their real reference objects\cite{Machprinciple}.

From above analysis, the new dynamical equation (9)(or (10)) is more universal than traditional Newton's second law, but now it is proved that this new dynamical equation is able to be derived totally under the framework of Newtonian mechanics. If both of them are correct, there seems to be some conflicts between them. In essence, this logical problem can be avoided.

In a more reasonable comprehension, the traditional Newton's second law should be restored into an empirical law summarized from a large number of classical mechanics experiments, namely a differential form of causal relationship between the new additionally exerted force(compared with a previous mechanical state) and the resulting relative acceleration under the premise of reference frame being fixed. In general, this empirical law can be expressed by the following forms,
\begin {eqnarray}
\Delta{\bf F}=m\Delta{\bf a},    &  or  &  d{\bf F}=md{\bf a}.
\end {eqnarray}
In history, this quasi-differential form is just the basis to determine the calculating formula of common forces, such as gravitational force, frictional force, elastic force and so on. Once forces are defined in specific cases, the dynamical causal relationship can be tested in other more general cases. The correct integral form of this quasi-differential causal relationship should be given by
\begin {eqnarray}
\frac{{\bf F}|_{p}}{m_{p}}-\frac{{\bf F}|_{O}}{m_{O}}={\bf a}|_{p-\Sigma}-{\bf a}|_{O-\Sigma}={\bf a}|_{p-O}.
\end {eqnarray}
However, the theoretical formula of Newton's second law is given by $ {\bf F}|_{p}=m_{p}{\bf a}|_{p-O}$.
Therefore, the new classical particle dynamics equation (9) has just corrected a drawback existing in the traditional formalism of Newton's mechanics, if we don't keep understanding the Newton's second law as mere a quasi-differential causal relationship for particle dynamics.

By making a comparison between the equation (9) and (1), the nature of inertial forces is revealed,
\begin {eqnarray}
{\bf f}|_{inertial}=-\frac{m_{p}}{m_{O}}{\bf F}|_{O}.
\end {eqnarray}
For any irrotational (with respect to the background of the universe) real reference frame, the nature of the inertial force is the real force acting on the reference object from the whole universe. In principle the inertial force can be all kinds of fundamental interactions, such as electromagnetic interaction, gravitational interaction, and so on. It is worth noting that such an interpretation of inertial forces is in conflict with Einstein's equivalence principle\cite{weinberg}. As it is well known, the concept of inertial force still exists in Einstein's theory of relativity, and Einstein's equivalence principle claims that the inertial force is equivalent to the gravitational force on all their physical effects, so the road map of Einstein's how to realize a general principle of dynamical relativity is outlined. However, we can see from the equation (9), which obviously is more natural and simple in physical picture. The so-called inertial force can actually be all kinds of common forces such as the friction force, traction force, gravitational force and so on. It must bring some influences on the physical picture of space-time since so far we know that only the gravitational force has the time dilation effect. In this sense, we have just found a counterexample of Einstein's equivalence principle. The deep influence on Einstein's general relativity brought about by such a new interpretation of inertial forces is deserved to be paid enough attentions.

A moderately general principle of relativity is essentially a practical requirement. On one hand, we are never able to know about the actual state of motion of our terrestrial reference frame where our observers exist. On the other hand, we are always able to determine the rotation of any reference frame with respect to cosmic spatial background by resorting to the galaxies far enough away since the cosmic spatial background is objective and motionless. In principle, the exact rotation of any practical reference frame with respect to the absolute background of space can be mathematically solved because the reference frame given in practice must interpret the dynamics for all objects in the universe, simultaneously and self-consistently. Therefore, what the practical observation really requires is that dynamical laws must keep form invariant to any reference frame which is irrotational with respect to the absolute background of space. Furthermore, in particle dynamics, the rotation phenomena can always be attributed to the relative motion between different particles. But for single particle, there is no concept of rotation. In other words, once any reference object is regarded as a particle, no problem of rotation exists for reference particle. Just as the problem of variable mass system under the framework of Newtonian mechanics, the variable mass phenomena should be attributed to the relative motion between different particles in the system of particles, so the fundamental particle dynamical equation is still ${\bf F}|_{p}=m_{p}{\bf a}|_{p-O}$. But for ${\bf F}|_{p}= \frac{d({\bf p}|_{p-O})}{dt}$, it actually can be generalized from the former equation when a system of particles is considered. In this sense, the problem of reference frames' rotation is essentially a mathematical problem. Ultimately, the problem of the rotation of reference frames can be separated from the problem of dynamical relativity.

Finally, the new particle dynamical equation (9) can be proved to be compatible with the existing theoretical structure of classical mechanics. If we adopt the new dynamical equation (9) as the fundamental particle dynamical equation in classical mechanics, there emerges a new question immediately. How should we deal with many theorems and deductions based on Newton's second law in classical mechanics? In fact, this problem can be easily solved without any unnaturalness. Because it doesn't matter what is the real counting range of ${\bf f}$ which gets involved in the derivation of these theorems and deductions. In other words, whether the force term ${\bf f}$ should be totally counted ${\bf f}={\bf F}|_{p}$, or relatively counted ${\bf f}={\bf F}|_{p}-\frac{m_{p}}{m_{O}}{\bf F}|_{O}$, does not affect the validity of the formulas of these theorems and deductions. Therefore, the new dynamical equation can be naturally and easily incorporated into existing theoretical structure of classical mechanics, and the only thing need to do is that all force terms appeared in these theorems and deductions should be understood as relatively counted forces between the particle under study and the reference object. Besides, as we have known, another formalism of classical mechanics is analytical mechanics. Under the least action principle of analytical mechanics, forces are depicted by potential energies. However, there is a premise that under the condition of reference frames being fixed, any work which is done by forces should be path independent, and completely determined by particles' initial and final states. Therefore, the real meaning of potential energy is a relatively counted energy which is defined directly based on the relative position between the particle under study and the reference object. In this sense, the spirit to relatively count forces in our new dynamical equation (9) is naturally in harmony with analytical mechanics.

\section{6 Physical Significance of the New Dynamical Equation}
First, the new dynamical equation (9) can somewhat improve the precision in the physical application. In principle, we should always make an approximation on the inertial reference frame before Newton's second law can be applied. Actually, such an approximation is made in theory, rather than in practical measurement. Therefore, if the contribution from the forces acting on the reference object can not be ignored, the error would be significant. By contrast, if we adopt the equation (9) as the new particle dynamical law, it at least has entirely solved the problem of the approximation on inertial reference frames. Since some research has indicated that the flat problem of galaxy rotation curve may be elegantly solved if the Newtonian dynamics is modified on the level of very small acceleration\cite{milgrom,ignatiev,gundlach}, it deserves to further investigate what kind of impact might be brought about by the application of the new dynamical equation (9).

Second, the new dynamical equation (9) illustrates that a more realistic general principle of relativity for particle dynamics can be realized in a very simple and natural approach which is obviously different from Einstein's view\cite{Bernard,Liuliao,weinberg}. Because the concept of inertial force still exists in Einstein's theory of relativity, and Einstein's equivalence principle further claims that the inertial force is equivalent to the gravitational force on all their physical effects. However, we can see from the equation (9), which obviously is more natural and simple in physical picture. The nature of the inertial force is the real force acting on the reference object. Therefore, the so-called inertial force can actually be all kinds of interactions such as the gravitational interaction, electromagnetic interaction and so on. It will bring some influence on the physical picture of space-time\cite{metric} since so far we know that only the gravitational interaction has the time dilation effect. In this sense, we have just found a counterexample of Einstein's equivalence principle.

Third, a logical derivation of the new dynamical equation (9) in Sec.3 strong suggests the existence of an absolute background of space. In fact, the background of space can be directly perceived. On the macroscopic scale, any empty space which we have seen is actually a part of the absolute background of space. For example, if an object is taken away from a certain place, the spatial region originally occupied by this object will not disappear with the removing of the object. The existence of this phenomenon partly reflects the existence of an absolute background of space. On the cosmological scale, the background of space is just the common background which reflects the motion of all galaxies in the universe. For instance, when any two adjacent galaxies continually moved away from each other, the empty space vacated between them is a highly approximated background of space. In the light of this new recognition, the concept of space should be further subdivided into two aspects. One is the relative length or duration of base units of space. And the other is the absolute background of space. The absolute background of space should be defined as the premise of the existence of the relative length or duration of the base unit of space, and the reference basis only on which any change of the length or duration of the base unit of space can be observed. Exactly speaking, the length or duration of the base unit of space should be regarded as a unit line segment which is cut from the absolute background of space.

\section{7 Conclusion}
In this paper we have discussed the formalism of particle dynamics under the framework of classical mechanics. The main motivation comes from a theoretical problem of Newton's second law, which is its application still depending the inertial frame of reference. Objectively speaking, Newton's second law is essentially an empirical law from ground based experiments which depicts the quantitative relation between the new additionally exerted force and the resulting relative acceleration. We must thereby distinguish such an empirical law from the traditional theoretical formula of Newton's second law. Consequently, a further reconstruction of particle dynamics based on this empirical law is required according to a moderately general principle of relativity. In precision, the new dynamical equation (9) is at the same level as traditional Newton's second law after the latter introduces a fictitious force by hand and approaches to its theoretical limits. In formalism, the new dynamical equation (9) can be directly applied in any reference frame without assuming an inertial reference frame and introducing a fictitious inertial force by hand. In concept, the new dynamical equation (9) removes the special status of inertial reference frames and inertial forces from its application, and presents a more concise physical picture based on reinterpreting all existing experiments for classical mechanics. Therefore, we suggest that the current theoretical formula of Newton's second law (1) should be replaced by the new dynamical equation (9) in the actual mechanics analysis.

{\bf{Acknowledgments:}} This work has been supported from the Nature Science Foundation of Zhejiang Province
under the grant number Y6110778. This research was also supported in part by the Project of Knowledge Innovation Program (PKIP) of Chinese Academy of Sciences, Grant No. KJCX2.YW.W10. This paper is dedicated to my grandfather Mr.Chen ChengDong.

\section{Appendix}
\section{RELATIVE BASE UNITS AND ABSOLUTE BACKGROUND IN THE PHYSICAL PICTURE OF SPACETIME}
\subsection{The conceptual survivability of absolute background under the proven physical logic in Einstein's theories of relativity}
The currently admitted physical theories of spacetime are Einstein's special theory of relativity and general theory of relativity. But if we reinvestigate the main physical logic in these two theories, we will find that both of them can substantially be understood as the change rules of the length or duration of base units of spacetime\cite{Sanhui,weinberg}. Just as its name implies, the length of base unit of space is the spatial span of the standard one meter, and the duration of base unit of time is the time span of the standard one second. In physical pictures of Einstein's theories of relativity, the concept of background of spacetime is not deliberately distinguished from the concept of the base units of spacetime. There is no concept of the background of spacetime in Einstein's theory. But in fact, the physics of the background of spacetime has been implicitly included in both Einstein's theories of relativity.

1), in the physical logic of special theory of relativity, every event is assumed to have an objective position in spacetime manifold when its coordinates are transformed between arbitrary two inertial reference frames. Otherwise, the Lorentz coordinate transformation cannot be obtained. Here the objective position of a physical event in spacetime manifold means that the event's occurring point in spacetime manifold doesn't change with the inertial reference frames. The existence of an objective position in spacetime manifold can actually be regarded as the reflection of the existence of an absolute background of spacetime\cite{Sanhui}.

2), in the physical picture of general theory of relativity, gravitational fields will result in a dilation effect for the duration of base unit of clocks and a contraction effect for the length of base unit of rulers. In other words, the span of base units defined in standard clock or standard ruler will be changed owing to the existence of a gravitational field. But theoretically, we should have a deeper picture for these physical effects. How can such a change in the spatial span of the base unit in standard rulers be embodied? There may be only one answer survivable. That is the existence of an absolute background of spacetime. Only when the base units of spacetime are compared with the absolute background of spacetime, the changes in their length or duration can be reflected and such an effect can be physical. More specifically, the base unit of standard ruler or stand clock is directly defined by the unit interval between two local intrinsic physical events which periodically occurs in specific objects, and the spatial span of the base unit of the standard ruler is just the line segment which is cut from the background of spacetime by the corresponding two local intrinsic events of the given unit interval.

Therefore at least in physical concepts, the existence of the background of spacetime can be compatible with the length or duration of base units of spacetime in a generalized physical picture based on Einstein's theories of relativity. Moreover, it is necessary to distinguish these two concepts, since the length or duration of base units of spacetime are relatively changeable according to Einstein's theory, but the background of spacetime must be absolute. Regarding the absolute background of spacetime, Natan Rosen has ever proposed a kind of bi-metric theories. He introduced an extra metric for flat space, in parallel to Einstein's curved metric, and both of them coexist in his theory\cite{Rosen}. The concept of flat space in Rosen's theory is a little close to here absolute background of spacetime, but they are different. Viewed from the side, Rosen's flat space is at least induced from his bi-metric theory to modify the general theory of relativity. But in our theory, there is always only one metric get involved. The spacetime metric normally describes the curve of spacetime or the change of length and duration of base units of spacetime under the existence of gravitation. And if all the matter in the universe is entirely absent, the spacetime metric must be reduced to be that of a flat Minkowski spacetime which substantially describes the absolute background of spacetime with the mathematically introduced base units of spacetime by observers.

\subsection{Physical concepts}
First of all, the length or duration of base units of spacetime and the background of spacetime are essentially the two aspects of spacetime, instead of two kinds of spacetimes. In physical concept, the absolute background of spacetime should be defined as the premise of the existence of relative length or duration of base units of spacetime and the reference basis only on which any change of the length or duration of base units of spacetime can be observed. In essence, the length or duration of base units of spacetime can be regarded as a unit line segment which is cut from the absolute background of spacetime. Taking a flat two-dimension plane for example, only after we define the length or duration of coordinate base units, a coordinate system is able to be painted on, and so we have a measurable concept of length for spatial spans on this two-dimension plane. But how far is the length of one meter of the standard ruler? To answer this question, the bottom board of this plane is indispensable. Without this bottom board serving as a foil to reflect, the length of one meter for the standard ruler will not make any sense. As an analogy, the background of spacetime is just equivalent to the bottom board of this two-dimension plane. If we ponder over it more deeply, the four-dimension background of spacetime may be imagined as a blank sheet of four-dimension paper. Originally, there is no coordinate on it. It is nothing but the observation that requires the introduction of coordinate base units. We can only define the base coordinate units by resorting to the local intrinsic events which periodically occur in specific objects, so the coordinate system is established.

Secondly, the base units of spacetime are physically defined by the unit intervals of local intrinsic events which occur in specific objects, on account of the requirement of measurements from observers. For instance, the second is the base unit of time in the International System of Units (SI). Since 1967, the second has been defined to be the duration of 9192631770 periods of the radiation corresponding to the transition between the two hyperfine levels of the ground state of the Caesium 133 atom\cite{BIPM}. Therefore in principle, the interval of spacetime between local intrinsic events is able to change, but the background of spacetime as the basis to reflect this change so must be homogeneous forever.

\subsection{The length or duration of the base units of spacetime}
Essentially, the base units of spacetime are directly defined by the unit intervals of local intrinsic events which periodically occur in specific objects. In this way, the length or duration of a spacetime base unit is described by the span of the line segment which is cut from the background of spacetime by the corresponding two local intrinsic events. There are long and short line segments. There are thus large and small base units of spacetime. Specifically, the duration of base unit of time can be denoted by the length of the line segment (or duration): $\overline{\Delta\tau(=1)}$, which is cut from the background of spacetime by the unit interval of local intrinsic events (${\Delta\tau=1}$).

\subsection{The reading number of the ruler and clock}
In contrast, the reading number of observers' clocks and rulers are substantially determined by the number of times for local intrinsic events which occur. Therefore, the reading number of clocks or rulers itself does not directly contain any information of spacetime base units. Since every physical event has its objective position in the background of spacetime, the change of a spacetime base unit can be determined by making a comparison of the fore-and-aft reading numbers of clocks or rulers as long as their corresponding line segments have the same length in the background of spacetime. The reading number of local intrinsic clock is just the time recorded by local observer, which can be denoted by $\Delta\tau$. Consequently, it should definitely be able to distinguish the reading number of local intrinsic clocks ($\Delta\tau$) and the length of the corresponding line segment cut from the background of time ($\overline{\Delta\tau}$), for every interval of events. In physical concepts, we should distinguish the local intrinsic clock and the observer's clock. The interval $dt$ or $dr$ appears in the curved invariant spacetime interval ($ds$) is usually the reading number of the observer's clock or observer's ruler.

\subsection{The changeability of the length or duration of physically defined base units of spacetime}
In special theory of relativity, a relative velocity between different reference frames will result in a dilation effect for the duration of clocks and a contraction effect for the length of rulers. It means that a spacetime base unit which is originally defined by the unit interval of the same kind of local intrinsic events may be different in length in the eyes of different observers. Nevertheless, in special theory of relativity both effects are simultaneously valid for each other of two observers, so it can be understood as an observational effect\cite{Sanhui}.

In general theory of relativity, a gravitational field will also result in a dilation effect for duration of clocks and a contraction effect for the length of rulers. Therefore, a spacetime base unit which is originally defined by the unit interval of the same kind of local intrinsic events will change in length or duration under the different gravitational field strength. In other words, there is a relative evolution which may exist for a local intrinsic clock. According to solar gravity tests, the geometric theory of gravity can be understood as a physical law for the change of the length or duration of physically defined base units of spacetime. As for the relation between the local intrinsic base units of spacetime and the observer's base units of spacetime, there is a really important assumption introduced. It is that the gravity causes the curvature of spacetime but in an infinitesimal neighborhood the spacetime should be asymptotically flat.

\subsection{The flatness, homogeneity and absoluteness for the background of spacetime}
First of all, it should be pointed out that any concept of flatness, homogeneity and absoluteness for any object should be defined by comparing with a more basic reference background. Therefore, if the background for the spacetime in whole universe has been set to be the most basic background, in the eyes of the observer, it should congenitally be regarded (or defined) to be flat and homogeneous. Because once it is not flat or homogenous, then such a conclusion must be made based on a more basic reference object. But as what we have just defined, the background for the spacetime in whole universe is set as the background at the most fundamental level. Therefore, it is always valid to say that the background of spacetime is flat and homogeneous. Similarly, we can always say that the background for the spacetime in whole universe is absolute. The reason is that we have defined the background for the spacetime in whole universe as the remained physical state after we have removed away all movable or evolvable objects from the current universe. Therefore, once the background is not absolute, it implies that this background must evolve with respect to a more basic reference object. However we have set the background for the spacetime in whole universe as the most fundamental reference. Consequently, it is also valid to say that the background for the spacetime in whole universe is absolute.
¡¡
\subsection{The preexistence and perpetuity for the background of spacetime}
There are many discussions about the creation of the universe in modern cosmology\cite{Grishchuk,Ferrara}. But incorporating above physical picture of spacetime, there one point which must be made clear is that, the so-called creation of the universe should be only limited to matter in our observable universe, instead of the background for the spacetime in whole universe. If the whole universe is really created from a thorough nothing, it means that such a creation doesn't require any premise or any precondition. Therefore, new universes would be created anytime and anywhere. This is not true. In this sense, the background of spacetime should preexist and last forever.

Besides, it is also meaningful to discuss the simultaneity in the background of time. In principle, the simultaneity in the background of time always exists according to a basic hypothesis that the background of time passes homogeneously. But an observable simultaneity should be artificially defined. For instance, if an observer wants to make clear the simultaneity between different spatial positions by means of the observation of physical phenomena, he has to resort to the number of times of local intrinsic events which occur on these spatial positions. In other words, the observable simultaneity should be determined by the coordinate values of spacetime manifold. Moreover, if we want to make a precise definition of the observable simultaneity, some physical interaction with invariant propagation speed may be required. For example in Einstein's special theory of relativity, this observable simultaneity is defined by the principle of the invariance of light speed, which is placed top priority. Therefore, an observable simultaneity is not always available for us in many cases. But for two events which occur on the same spatial position, we will definitely be able to distinguish the time order of the occurrence, so we always can retain the concept of simultaneity for the same spatial position. Therefore, the simultaneity in the background of time always exists objectively. But the directly observable simultaneity for observers must be defined by resorting to specific physical phenomena.

\end{document}